\documentclass[sigconf,nonacm]{acmart}

\AtBeginDocument{%
  \providecommand\BibTeX{{%
    \normalfont B\kern-0.5em{\scshape i\kern-0.25em b}\kern-0.8em\TeX}}}

\usepackage{graphicx} 
\usepackage{xspace}
\usepackage{xcolor}
\usepackage{hyperref}
\usepackage{amsmath}
\usepackage[linesnumbered,ruled,vlined]{algorithm2e}
\usepackage{multirow}
\usepackage{amsmath}
\usepackage{cleveref}
\usepackage[frozencache,cachedir=.]{minted} 
\usepackage[most]{tcolorbox}
\usepackage{siunitx}
\usepackage{enumitem}
\setlist{leftmargin=7mm}

\newcommand{\ie}{\textit{i.e.},~}
\newcommand{\eg}{\textit{e.g.},~}
\newcommand{\projName}{\textsc{SWEzze}\xspace}
\begin{document}

\title{Compressing Code Context for LLM-based Issue Resolution}

\author{Haoxiang Jia}
\email{haoxiangjia@stu.pku.edu.cn}
\affiliation{%
  \institution{Peking University}
  \city{Beijing}
  \country{China}
}

\author{Earl T. Barr}
\email{e.barr@ucl.ac.uk}
\affiliation{%
  \institution{University College London}
  \city{London}
  \country{UK}
}

\author{Sergey Mechtaev}
\authornote{Sergey Mechtaev is the corresponding author.}
\email{mechtaev@pku.edu.cn}
\affiliation{%
  \institution{Peking University}
  \city{Beijing}
  \country{China}
}

\begin{abstract}
Large Language Models (LLMs) are now capable of resolving real-world GitHub issues. However, current approaches overapproximate the code context and suffer from two compounding problems: the prohibitive cost of processing massive inputs, and low effectiveness as noise floods the context window and distracts the model from the bug-fixing signal. Existing compression techniques fail to resolve this tension: generic compressors compromise the semantic integrity of code, while code-specific tools lack awareness of code structure and task context to preserve essential patch ingredients. To address this, we propose a novel framework consisting of two components. First, Oracle-guided Code Distillation (OCD), a context distillation algorithm that combines genetic search and delta debugging to systematically reduce code contexts to their minimal sufficient subsequence --- retaining only the ingredients required for a successful fix. We use this distilled data to fine-tune \projName, a lightweight model that learns to compress code context at inference time, filtering noise and combating distraction while preserving fix ingredients. Evaluated on SWE-bench Verified across three frontier LLMs, \projName maintains a stable compression rate of about 6$\times$ across models, reduces the total token budget by 51.8\%--71.3\% relative to the uncompressed setting, improves issue resolution rates by 5.0\%--9.2\%, and delivers the best overall balance among effectiveness, compression ratio, and latency compared with state-of-the-art context compression baselines.
\end{abstract}


\maketitle

\section{Introduction}

Issue resolution by Large Language Models (LLMs) has progressed from academic prototypes to practical adoption in industry. Systems such as Agentless~\cite{xia2024agentless} and SWE-Agent~\cite{DBLP:conf/nips/YangJWLYNP24} can localize faults, synthesize patches, and resolve issues across real-world repositories. Yet scaling these approaches reveals a fundamental bottleneck: cost and effectively utilizing code context. Unlike many natural-language tasks, issue resolution requires reasoning over context scattered across multiple files --- definitions, call chains, class hierarchies, modules, and interfaces --- making context construction a critical component of the resolution pipeline~\cite{li2026contextbench}.
  
To extract relevant code context, existing approaches rely on heuristic retrieval or Retrieval-Augmented Generation (RAG)~\cite{lewis2020retrieval} to rank files according to lexical similarity, embeddings, or shallow structural signals. To avoid missing crucial dependencies, these methods typically return top-$k$ files or functions that collectively span hundreds of lines to capture only a handful of relevant segments. This leads to context overapproximation, where the true bug-fixing signal is diluted by large amounts of irrelevant code. The impact is twofold. First, it imposes a direct computational inference cost: inference cost scales roughly with the size of the provided context~\cite{DBLP:conf/nips/VaswaniSPUJGKP17}. Second, excessive context degrades effectiveness, as irrelevant code floods the context window~\cite{liu2024lost} and distracts the model from the bug-fixing signal~\cite{shi2023large}.

A promising mitigation to this problem is to compress retrieved contexts: distilling the overapproximate output of retrieval into a concise form that retains only the essential information, and use it as the input to the downstream issue-resolving LLM. However, existing compression strategies~\cite{li2025prompt} are ineffective for patch generation. Generic compressors~\cite{jiang2024longllmlingua} that treat code as natural language tend to disrupt program structure and break semantic links (e.g., definition–use relations and type constraints) that are often required to construct a correct patch. Conversely, code-specific heuristic pruners~\cite{shi2025longcodezip,wang2026swe} select context based on its similarity or relatedness to the issue description or the buggy code, but issue resolution often depends on fix ingredients~\cite{martinez2014fix,liu2018lsrepair,white2019sorting}, \ie code elements such as variables, expressions, and types that are needed to construct a correct patch, that may not bear obvious lexical or structural resemblance to the bug itself, causing these compressors to discard the necessary ingredients.

To address this problem, we propose a framework that distills high-quality training data from a computationally expensive oracle to train a lightweight, inference-ready model. We first employ Oracle-guided Code Distillation (OCD) to systematically reduce an initial context   to a minimal sufficient context (MSC) --- a one-minimal subsequence where every retained fragment is functionally necessary for a successful repair. To produce these MSCs, OCD uses a Genetic Algorithm (GA) to discover patch-enabling configurations, which Hierarchical Delta Debugging (HDD)~\cite{misherghi2006hdd} then minimizes. These distilled examples from the SWESmith dataset~\cite{yang2025swe} serve as the training signal for \projName, a cross-encoder fine-tuned from Qwen3-Reranker-0.6B~\cite{zhang2025qwen3} via parameter-efficient LoRA adaptation~\cite{DBLP:journals/corr/abs-2106-09685}. By learning to map initial code context $C_\mathit{init}$ to its corresponding MSC ($\hat{C}$), \projName captures the structured dependency patterns that distinguish fix ingredients~\cite{martinez2014fix} from incidentally related code. At inference time, \projName operates without oracle access, scoring segments and assembling compressed context through budget-aware greedy selection.

\begin{figure*}[t]
  \begin{center}
    \includegraphics[width=0.95\textwidth]{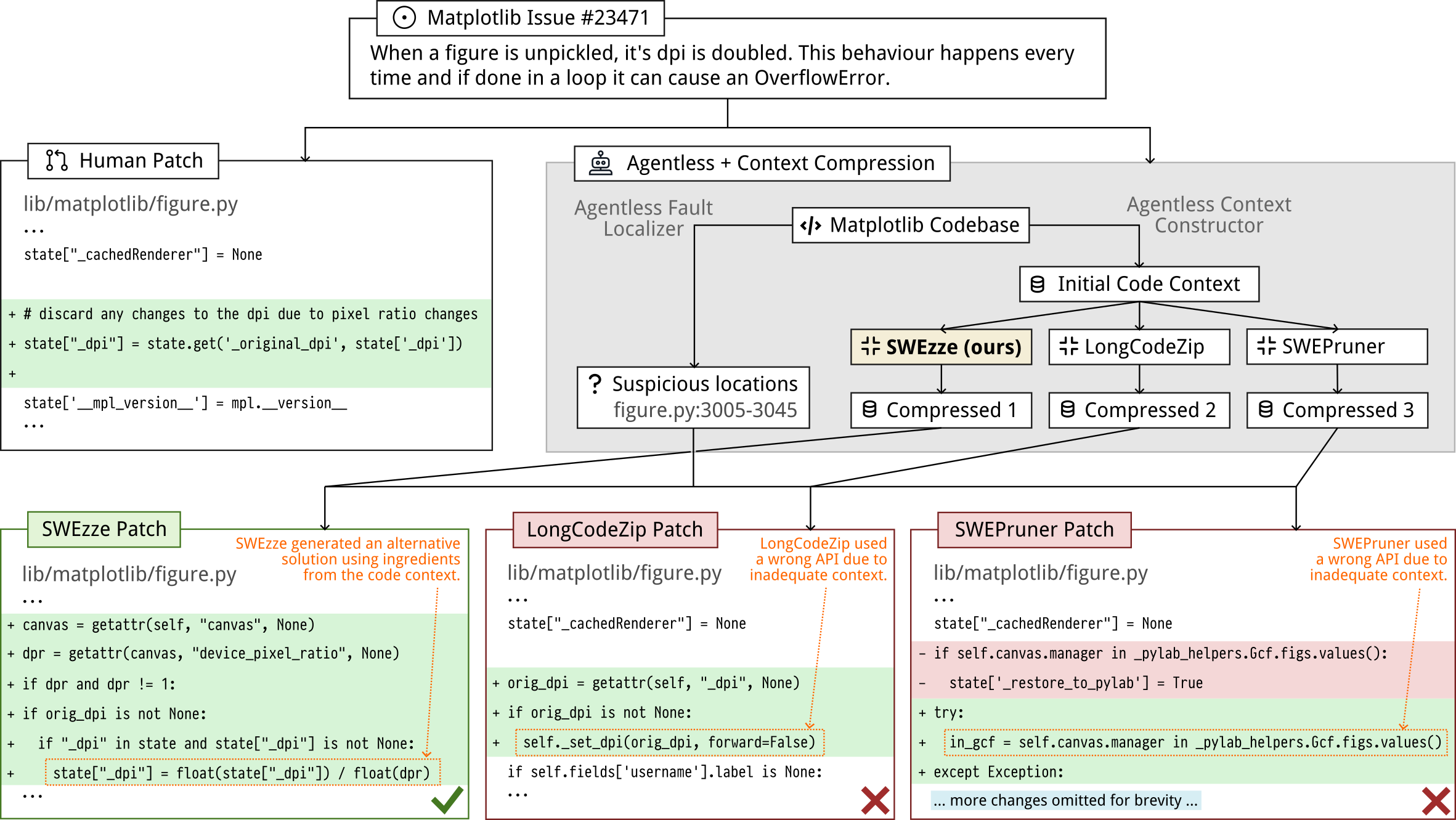}
  \end{center} 
  \caption{Agentless + \projName generates a correct patch because \projName retains the fix ingredients in the compressed context. In contrast, previous compressors (LongCodeZip and SWE-Pruner) degrade Agentless performance by removing these ingredients.\label{fig:resolution_workflow}}
\end{figure*}

We evaluate \projName within the Agentless workflow~\cite{xia2024agentless} on SWE-bench Verified across the GPT-5.2, DeepSeek-V3.2, and Qwen3-Coder-Next frontier models, and compare with state-of-the-art compressors. \projName achieves the best end-to-end issue resolution performance and the best balance among repair effectiveness, compression strength, and runtime overhead. Across all three models, it maintains a stable compression rate of about 6$\times$, reduces the total token budget by 51.8\%--71.3\% relative to the uncompressed setting, and improves issue resolution rates by 5.0\%--9.2\%. Compared with prior compressors, \projName not only achieves higher overall resolution rates, but also covers the broadest set of solvable instances, reaching 93.8\%--99.2\% of the union of all cases resolved by any baseline. Our results show that for repository-level issue resolution compression is more effective when it preserves repair-sufficient context, rather than maximizing token reduction or similarity.

In summary, the paper makes the following contributions:
\begin{itemize}
\item Oracle-guided Code Distillation, a novel framework for distilling minimal sufficient context for LLM-based issue resolution.
\item \projName, a novel code context compression model that retains fix ingredients while removing distracting noise.
\item Evaluation on SWE-Bench Verified that demonstrates that \projName significantly reduces the cost of LLM-based issue resolution while at the same time improving success rate.
\end{itemize}

All code, data, and reproduction scripts are available at \url{https://zenodo.org/records/19248411}.

\section{Motivating Example}

\begin{figure}[t]
  \begin{center}
    \includegraphics[width=\columnwidth]{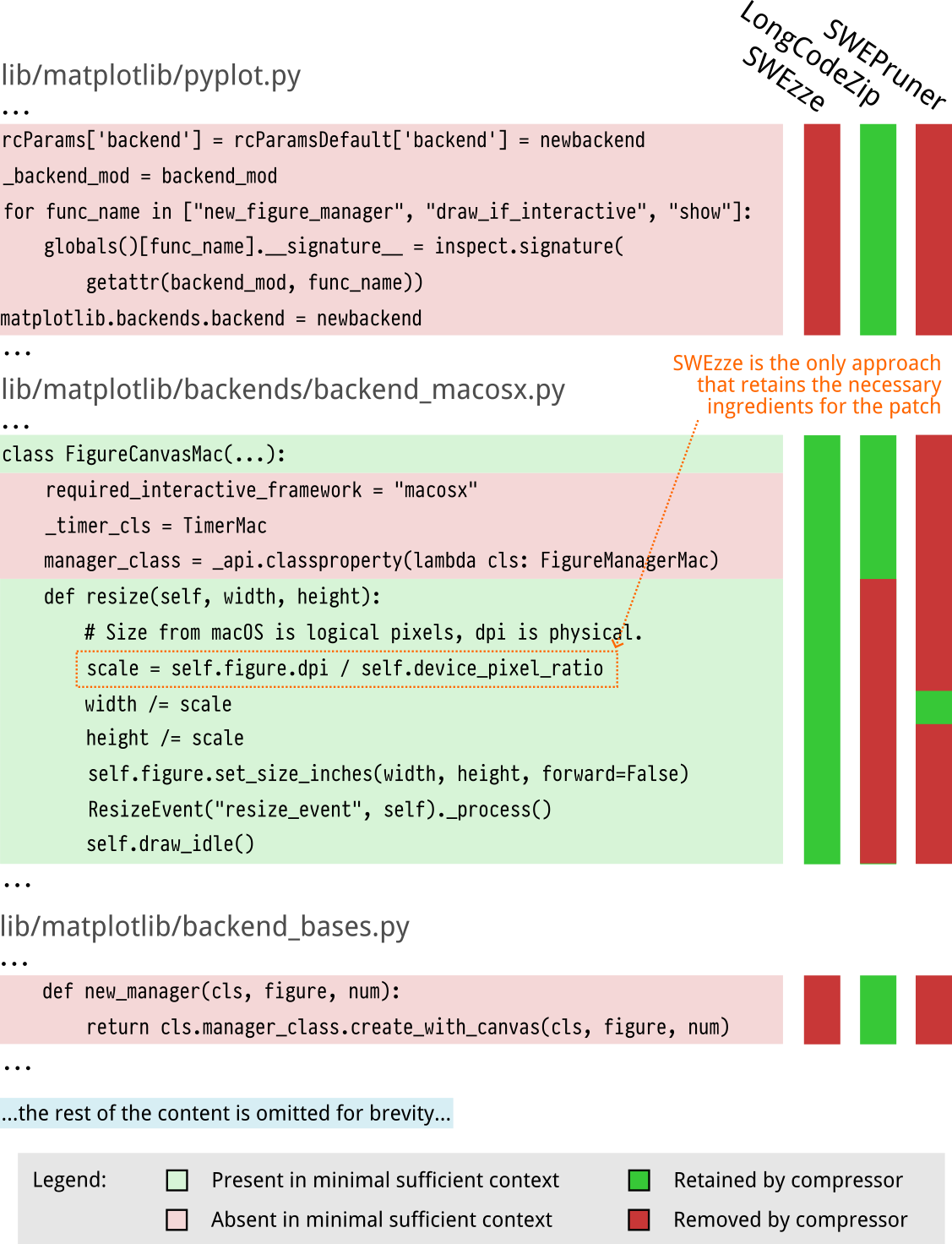}
  \end{center} 
  \caption{A fragment of Agentless' code context for resolving the issue in \Cref{fig:resolution_workflow}. \projName's context has the higher correlation with a minimal sufficient context computed via delta-debugging, and the only one that contains the ingredients to correctly recompute DPI to address the issue.\label{fig:context_comparison}}
    \vspace{-4mm}
\end{figure}

We illustrate \projName's ability to identify and retain fix ingredients for issue resolution using a case study from the widely-used Matplotlib library~\cite{matplotlib}, extracted from SWE-Bench~\cite{jimenez2023swe}. This issue, shown in \Cref{fig:resolution_workflow} (top), reports a bug in DPI calculation on Mac OS. The human patch for this issue (top left) simply inserts a state update statement inside the file \mintinline{Python}`figure.py`, using the API call \mintinline{Python}`state.get('_original_dpi', state['_dpi'])`.
Although the fix itself is small, it is representative of real-world issue resolution: constructing the patch requires understanding a project-specific API that is unlikely to appear in the issue description, making it essential for the compressor to retain the relevant context.

\projName compresses prompts by pruning their code segments to an issue's fix ingredients~\cite{martinez2014fix}, the minimal set of semantic dependencies needed to resolve that issue.
We designed \projName to be model- and harness-agnostic, ensuring compatibility with any downstream repair workflow. 
To illustrate its operation, we integrate \projName into Agentless~\cite{xia2024agentless}, a  pipeline combining fault localization and context construction that is widely used to evaluate frontier LLMs on issue resolution~\cite{deng2025swe}. 
The ``Agentless + Context Compression'' module (\Cref{fig:resolution_workflow}, top right) depicts Agentless localizing a bug and forwarding the initial code context to three compressors: \projName and two established baselines.

The leaf nodes of \Cref{fig:resolution_workflow} present the repair patches generated by the frontier model when provided the suspicious locations and the compressed contexts from each tool. As you can see, the resulting patches diverge significantly. When using the contexts from LongCodeZip~\cite{shi2025longcodezip} and SWE-Pruner~\cite{wang2026swe}, the model produces patches that do not resolve the issue. In contrast,  \projName's context preserves the fix ingredients, enabling the model to generate a patch that matches the logic of the human reference. Agentless' success resolving this issue depends critically on the choice of context compressor: it resolves the issue only when using \projName, because the baseline compressors discarded crucial, fix-relevant information.

To understand this difference, we compare the compressed contexts in \Cref{fig:context_comparison}. This figure shows a fragment of the initial code context $\mathcal{C}_\mathrm{init}$, where the background color indicates whether a segment appears in the fix ingredients $\hat{\mathcal{C}}$ identified via delta debugging~\cite{zeller2002simplifying}: light green denotes retained segments, light red denotes removed ones. Notably, although $\hat{\mathcal{C}}$ does not contain any references to \mintinline{Python}`'_original_dpi'` used in the human patch, it contains the statement

\mintinline{Python}`scale = self.figure.dpi / self.device_pixel_ratio`,

\noindent which reveals how the scale is computed. With this fragment in its context, the downstream LLM correctly recomputes DPI as shown in the patch generated with \projName in \Cref{fig:resolution_workflow}, bottom left. \projName is the only compressor that retains this fragment: LongCodeZip selected irrelevant fragments, while SWE-Pruner retained only a single line from this function, which proved insufficient for the model to construct a fix.

This performance reflects \projName's core design: an oracle that protects fix ingredients and a search strategy (GA and HDD) that maximizes our oracle budget. While baseline compressors rely on static heuristics, \projName's search-based OCD framework identifies and retains the functional dependencies required for a successful repair. Consequently, \projName achieves a BERTScore~\cite{DBLP:conf/iclr/ZhangKWWA20} of 0.44 against the minimal context $\hat{C}$, indicating high semantic overlap with the fix-relevant fragments, more than doubling the fidelity of LongCodeZip (0.20) and SWE-Pruner (0.00). This approach allows \projName to preserve critical components like the \mintinline{Python}{'resize'} method, which the baselines systematically strip away.


\section{Oracle-Guided Context Distillation}
\label{sec:ocd}

To train a compression model that preserves fix ingredients, we need training data in which these ingredients are distilled to their essence. We operationalize this through the notion of \emph{minimal sufficient context}. Given an issue description $\mathcal{I}$, a fault location $\mathcal{F}$, and an initial over-approximate code context $\mathcal{C}_\mathrm{init}$ retrieved from a repository, the goal is to identify a subset $\hat{\mathcal{C}} \subseteq \mathcal{C}_\mathrm{init}$ such that $|\hat{\mathcal{C}}| \ll |\mathcal{C}_\mathrm{init}|$. We call $\hat{\mathcal{C}}$ a \textit{sufficient code context} if an LLM can use $\mathcal{I}$, $\mathcal{F}$, and $\hat{\mathcal{C}}$ to generate a patch $\mathcal{P}$ that passes the validation suite $\mathcal{T}$. Sufficient contexts are generally not unique: multiple distinct subsets of $\mathcal{C}_\mathrm{init}$ may independently satisfy sufficiency while being incomparable under set inclusion (\eg two alternative helper functions may each enable a correct patch, yet neither subsumes the other). Our goal is to find any sufficient context that is also minimal.

Finding a globally minimal (minimum-cardinality) sufficient context would require enumerating all $2^{|\mathcal{C}_\mathrm{init}|}$ subsets---computationally intractable for realistic codebases. Following delta debugging~\cite{zeller2002simplifying}, we instead target \textit{1-minimality}, a local minimality criterion that is efficiently computable yet provides strong compression guarantees:

\begin{definition}[Minimal Sufficient Context]
A sufficient code context $\hat{\mathcal{C}} \subseteq \mathcal{C}_\mathrm{init}$ is \textit{minimal} if removing any single element from $\hat{\mathcal{C}}$ causes it to become insufficient---\ie for every element $e \in \hat{\mathcal{C}}$, $\hat{\mathcal{C}} \setminus \{e\}$ does not enable the LLM to generate a patch that passes $\mathcal{T}$.
\end{definition}

The 1-minimality ensures that every retained element is individually necessary, producing clean training signal without redundant context. At the same time, it is tractable: verifying 1-minimality requires at most $|\hat{\mathcal{C}}|$ oracle calls per element, unlike the exponential cost of global minimality.

The goal of \emph{Oracle-Guided Context Distillation (OCD)} is to produce high-quality training examples that map $\mathcal{C}_\mathrm{init}$ to a 1-minimal sufficient subset $\hat{\mathcal{C}}$. Unlike prior compression methods that rely on perplexity or embedding similarity, we define sufficiency through a functional criterion. We formulate context distillation as a search problem guided by an execution oracle that evaluates candidate contexts by invoking the repair model and validating generated patches. Our approach comprises three components: (1) a hierarchical search space representation that ensures syntactic validity (\S\ref{sec:search_space}); (2) priority scoring that leverages training-time information to guide search (\S\ref{sec:priority}); and (3) a two-phase search strategy combining genetic algorithm and delta debugging (\S\ref{sec:search_strategy}).

\subsection{Search Space Representation}
\label{sec:search_space}

The effectiveness of search for minimal context depends on how the search space is structured. Token-level representation, while flexible, frequently results in syntactically invalid code that may confuse the model. To ensure that all candidate context subsequences remain well-formed, we impose a hierarchical structure on the search space that respects the syntactic rules.

We decompose each source file $F \in \mathcal{C}_\mathrm{init}$ into a tree of nested structural units. Throughout this paper, we use the term ``structural unit'' (or simply ``unit'') to refer to any element at these three hierarchical levels; the hierarchy consists of three levels:

\begin{itemize}
    \item \textbf{File Level:} File-level decisions determine whether any content from a source file appears in the compressed context.
    \item \textbf{Function Level:} Within a file, we view function and method definitions as independent units. This level captures the organizational structure of procedural and object-oriented code.
    \item \textbf{Block Level:} Within a function, we partition statements into contiguous blocks. Each block represents a sequence of statements that can be independently included or excluded.
\end{itemize}


We refer to the leaf-level units produced by the hierarchical decomposition as \textit{code segments} (or simply \textit{segments}). Concretely, a segment is one of five AST-derived types:
\begin{itemize}
    \item \textbf{method}: a method definition bound to a class;
    \item \textbf{function}: a top-level or nested function definition;
    \item \textbf{class\_header}: the signature and class-level statements of a class (excluding its method bodies);
    \item \textbf{block}: a contiguous sequence of statements inside a function body that does not constitute a standalone definition;
    \item \textbf{file}: a top-level fragment that is outside any class or function.
\end{itemize}
Segments are the atomic scoring units used by \projName and the granularity at which inclusion/exclusion decisions are made during both data construction and inference.

Rather than silently removing code units, we replace omitted content with placeholders. When a unit $u$ is excluded, we substitute it with a template $\tau_u$ that indicates the location and magnitude of the omission (e.g., \texttt{\# ... N lines omitted}, where $N$ denotes the number of source code lines in the original unit). This design preserves the code structure, allowing the LLM to recognize that additional context exists even if not explicitly provided.

\subsection{Priority-Guided Search}
\label{sec:priority}

The search space of context subsequences is exponential in the number of structural units. To make search tractable, we assign priority scores that guide the search towards promising regions. For each structural unit $u$, we compute a priority score $\Phi(u)$ that estimates its relevance to successful patch generation. The score aggregates three categories of signals:

\textbf{Patch Overlap.} Let $\mathcal{F}_\mathrm{patch}$ denote the set of files modified by the ground-truth patch $\mathcal{P}_\mathrm{gold}$. Units that belong to $\mathcal{F}_\mathrm{patch}$ receive elevated priority, as the developer's patch serves as direct evidence of which files require modification. We capture this signal using an indicator function $\mathbb{I}(u \in \mathcal{F}_\mathrm{patch})$ that returns 1 if unit $u$ belongs to these patch-relevant files and 0 otherwise.

\textbf{Test Coverage.} We execute the failing tests and record the set of covered lines $\mathcal{L}_\mathrm{cov}(u)$ within each unit $u$. Units containing more covered lines are prioritized, as test coverage indicates dynamic relevance to the failure. To prevent bias from test cases that exercise large portions of the codebase, we apply logarithmic dampening to the coverage count.

\textbf{Symbol Overlap.} We extract identifiers (variable, function, class names) from $\mathcal{P}_\mathrm{gold}$ and define $\text{SymScore}(u, \mathcal{P}_\mathrm{gold})$ as a function measuring lexical overlap between these identifiers and the content of unit $u$. This heuristic captures semantic dependencies: if a patch references a particular function or variable, the code defining that symbol is likely necessary for correct patch generation.

The composite priority score combines these three signals:
\begin{align*}
  \Phi(u) &= w_p \cdot \mathbb{I}(u \in \mathcal{F}_\mathrm{patch}) + w_c \cdot \log(1 + |\mathcal{L}_\mathrm{cov}(u)|) \\
  &+ w_s \cdot \text{SymScore}(u, \mathcal{P}_\mathrm{gold})
\end{align*}
where $w_p$, $w_c$, and $w_s$ are configurable weights controlling the relative importance of patch overlap, test coverage, and symbol overlap signals, respectively.


\subsection{Two-Phase Search Strategy}
\label{sec:search_strategy}

Having defined the search space and priority scoring, we now describe the search algorithm itself. The initial context $\mathcal{C}_\mathrm{init}$ may or may not enable successful patch generation --- failure can stem from inherent task difficulty or from distracting elements that mislead the model. Therefore, we first must identify a subset of $\mathcal{C}_\mathrm{init}$ that enables successful patch generation, before minimizing it.

We address this problem with a two-phase pipeline (Algorithm~\ref{alg:ocd}): a Genetic Algorithm (GA) identifies resolution-enabling subsequences solution, and Hierarchical Delta Debugging (HDD) minimizes them. The two phases are complementary: GA handles the combinatorial challenge of finding a sufficient subset among exponentially many candidates, while HDD ensures the result contains no redundant units. Priority scores $\Phi$ guide both phases---biasing GA initialization and fitness toward high-priority configurations, and guiding HDD to remove low-priority units first.

\begin{algorithm}[t]
\caption{Oracle-Guided Context Distillation}
\label{alg:ocd}
\KwIn{Context $\mathcal{C}_\mathrm{init}$, oracle $\mathcal{O}$, priority $\Phi$}
\KwOut{Minimal sufficient context $\hat{\mathcal{C}}$}
\tcc{Hierarchical Decomposition (\S\ref{sec:search_space})}
$\mathcal{U} \leftarrow \textsc{Decompose}(\mathcal{C}_\mathrm{init})$
  \tcp*{file $\supset$ function $\supset$ block}
$\Phi(u) \leftarrow \textsc{PriorityScore}(u)\;\forall u \in \mathcal{U}$
  \tcp*{\S\ref{sec:priority}}
\BlankLine
\tcc{Phase~I: Genetic Algorithm Search (\S\ref{sec:search_strategy})}
$\hat{\mathcal{C}}_{\mathrm{GA}} \leftarrow \textsc{GeneticSearch}(\mathcal{U},\, \Phi,\, \mathcal{O})$\;
\lIf{no passing subset found}{\Return $\mathcal{C}_\mathrm{init}$}
\BlankLine
\tcc{Phase~II: Hierarchical Delta Debugging (\S\ref{sec:search_strategy})}
$\hat{\mathcal{C}} \leftarrow \hat{\mathcal{C}}_{\mathrm{GA}}$\;
\ForEach{level $\ell \in \{\textit{file},\, \textit{function},\, \textit{block}\}$}{
    $\hat{\mathcal{C}} \leftarrow \textsc{ddmin}(\hat{\mathcal{C}},\; \ell,\; \mathcal{O},\; \Phi)$
      \tcp*{remove redundant units}
}
\Return $\hat{\mathcal{C}}$\;
\end{algorithm}

\subsubsection{Phase I: Genetic Search}
We employ a Genetic Algorithm (GA) to explore the space of context configurations. GA's population-based search evolves configurations toward promising regions of the search space. We represent each candidate context as a binary genome $\mathbf{g} = (g_1, g_2, \ldots, g_n)$, where each gene $g_i \in \{0, 1\}$ corresponds to a structural unit $u_i$ (file or function). A gene value of 1 indicates inclusion, while 0 indicates exclusion. To maintain syntactic validity, we enforce upward consistency: if a function gene is active, its parent file gene must also be active. This constraint is enforced after each genetic operation through a repair step. The fitness function is designed to identify passing solutions while guiding failed configurations toward promising regions. For failed configurations, the fitness function is the sum of retained code segment priority scores which guides the search toward larger, high-priority contexts more likely to contain patch-critical information.

We employ operators adapted for hierarchical code structure:
\begin{itemize}
    \item \textbf{Initialization:} The initial population is seeded with high-probability configurations derived from priority scores (\eg context restricted to gold-patch files). Remaining slots are filled with randomly generated individuals whose inclusion probability is biased by priority.
    \item \textbf{Selection:} Tournament selection with elitism preserves the top 20\% of individuals, ensuring good solutions are not lost while maintaining population diversity.
    \item \textbf{Crossover:} Hierarchical crossover operates at file granularity, swapping entire file subtrees between parents. This preserves the internal coherence of file structures and respects the natural modularity of code.
    \item \textbf{Mutation:} Standard bit-flip mutation is applied with a configurable rate, followed by a repair step that enforces upward consistency and prevents degenerate empty individuals.
\end{itemize}

GA stops immediately upon discovering any resolution-enabling subsequence solution rather than continuing to find better ones.


\subsubsection{Phase II: Hierarchical Delta Debugging}
Once a resolution-enabling context subsequence is discovered, we apply Hierarchical Delta Debugging (HDD)~\cite{misherghi2006hdd} to systematically minimize it. HDD extends the classical delta debugging algorithm by exploiting the hierarchical structure of our search space, ensuring the final context is locally minimal. It operates in three sequential passes, each targeting a different level of the hierarchy:
\begin{enumerate}
    \item \textbf{File-Level Reduction:} We attempt to remove entire files from the context. Files that can be removed without breaking sufficiency are permanently excluded.
    \item \textbf{Function-Level Reduction:} Within each retained file, we attempt to remove individual functions, identifying functions that were retrieved but are not actually required for the fix.
    \item \textbf{Block-Level Reduction:} Within each retained function, we attempt to remove statement blocks, eliminating unnecessary implementation details while preserving essential logic.
\end{enumerate}

At each level, we apply the \texttt{ddmin} algorithm~\cite{zeller2002simplifying} adapted for our setting. Given a set of currently retained units $U$, \texttt{ddmin} iteratively partitions $U$ and tests whether smaller subsets maintain sufficiency. The algorithm terminates when removing any single remaining unit causes the context to become insufficient, achieving \textit{1-minimality}. We integrate priority scores by sorting units in ascending priority order before partitioning. This causes the algorithm to preferentially attempt removing low-priority units first, accelerating convergence by front-loading the removal of redundant content.

\subsubsection{Candidate Evaluation}

Each candidate context evaluation requires invoking the repair model to generate patches and executing the test suite to validate correctness. This process is computationally expensive, making evaluation efficiency critical for scalable data construction. To reduce variance from the stochastic nature of LLM generation, we generate multiple patch candidates per evaluation and apply majority voting. A context is deemed sufficient if at least half of the generated patches pass all tests, providing more reliable sufficiency judgments.



\subsection{OCD on SWE-Smith}
\label{sec:characteristics_of_distilled_data}

To prepare training data for our compression model, we applied OCD to SWE-Smith~\cite{yang2025swe}, a large dataset of 50K issues sourced from 128 GitHub repositories, which does not intersect with our testing set. From this dataset, we uniformly sampled 10K issues, and distilled Agentless' contexts for these with OCD by using Qwen3-Coder-30b to evaluate context sufficiency. 

\begin{table}
\caption{Overview of the constructed training dataset based on SWE-Smith of minimal sufficient contexts.}
\label{tab:dataset_overview}
\vspace{-1mm}
\begin{tabular}{lr}
\toprule
Metric & Value \\
\midrule
Unique repositories & 41 \\
Total instances (bugs) & 3,157 \\
Total code segments & 156,545 \\
Avg. segments per instance & 49.6 \\
Positive segments (relevant) & 13,102 \\
Negative segments (irrelevant) & 143,443 \\
Relevance density & 8.4\% \\
\bottomrule
\end{tabular}
\vspace{-1mm}
\end{table}

\begin{figure}[t]
  \centering
  \includegraphics[width=\linewidth]{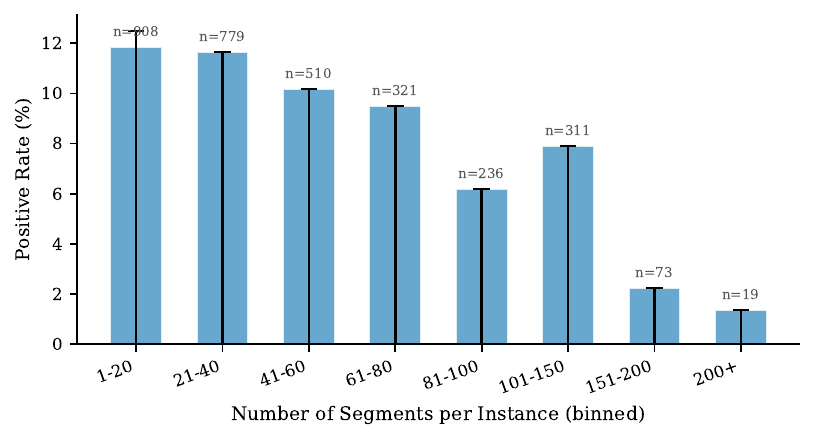}
  \vspace{-4mm}
  \caption{Relevance density vs.\ context size. Relevance density monotonically decreases with the increase of context size, confirming that broader retrieval introduces proportionally more noise. Error bars denote interquartile ranges.}
  \label{fig:positive_rate_vs_size}
    \vspace{-1em}
\end{figure}

Before fine-tuning a compressor, we analyzed the data that OCD produced from three dimensions: compression effectiveness, segment-level relevance, and context sparsity. Table~\ref{tab:dataset_overview} summarizes the distilled corpus. Across 3,157 instances from 41 Python repositories, for which a sufficient context was found, the AST segmenter produces 156,545 labeled segments, or 49.6 segments per instance on average. Only 8.4\% of segments are relevant to the fix, we call this ratio \textbf{relevance density}; on average, only 4.15 out of 49.6 segments per instance are necessary for patch generation. This severe class imbalance shows why a compressor cannot be trained with a naive uniform loss: the easy all-negative decision rule is statistically attractive but useless for repair. Figure~\ref{fig:positive_rate_vs_size} reveals that the fraction of relevant segments decreases monotonically with context size: instances with small context (1-20 segments) have a relevance density above 11\%, while those with 200+ segments drop below 2\%. At the segment level, method-level units constitute 78.0\% of all segments but have a relevance density of only 7.5\%, making method selection the primary bottleneck of context compression.

We note that HDD is required to produce minimal sufficient contexts by definition. To understand the impact of GA, we conducted an ablation --- when not using the genetic algorithm, the number of successfully minimized instances reduces by 52.7\%.

\begin{figure}[t]
  \centering
  \includegraphics[width=\linewidth]{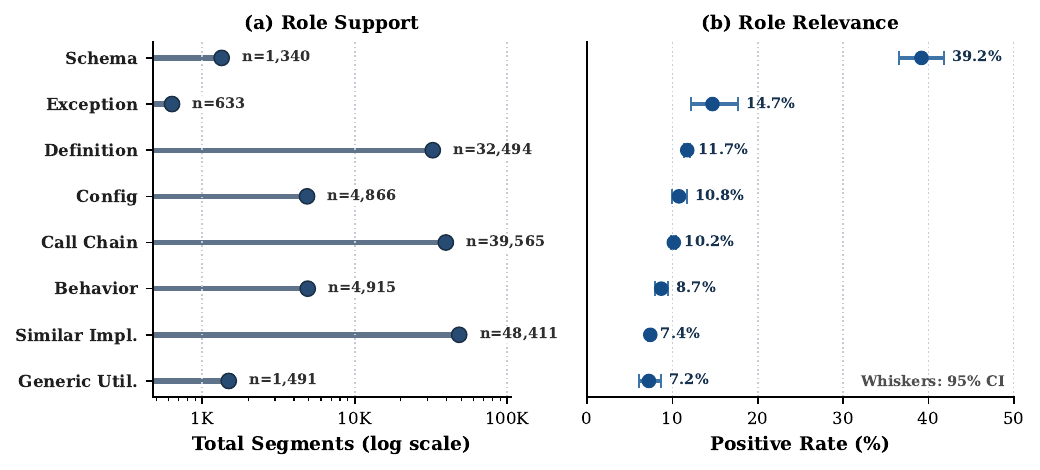}
  \caption{Relevance density per semantic role.
    Schema segments achieve the highest relevance density (39.2\%),
    while Generic Utility segments represent the strongest noise source (7.2\%).
    The wide variance across roles motivates role-aware sample weighting
    in \projName training.}
  \label{fig:semantic_role}
    \vspace{-1em}
\end{figure}

To understand why certain segments are relevant, we classify each segment into semantic roles based on its structural relationship to the issue description and fault location (Figure~\ref{fig:semantic_role}). The \textbf{Schema} role (data classes and type definitions) achieves the highest relevance density at \textbf{39.2\%}, reflecting the prevalence of type-mismatch and field-naming bugs in SWE-bench. \textbf{Definition} segments (11.7\%) and \textbf{Call Chain} segments (10.2\%) confirm that structural proximity to the fault in the call graph is a strong predictor of relevance. In contrast, \textbf{Generic Utility} segments exhibit the lowest relevance density (7.2\%), representing recall noise from broad file-level retrieval. The five-fold variance in relevance density across roles---from 7.2\% to 39.2\%---has a practical implication: a uniform loss function would under-weight rare but highly informative schema and definition segments while over-penalizing the model for retaining generic utilities. This observation motivates the role-aware sample weighting strategy adopted in \projName's training, where per-sample weights are modulated by semantic role to reflect the differential informativeness of each segment category.

\section{\projName Compression Model}
\label{sec:swezze}

Distillation in \Cref{sec:ocd} yields training examples that map an over-approximate context to a minimal sufficient subset. Formally, it produces a corpus $\mathcal{D}$ contains tuples of the form $(\mathcal{I}, \mathcal{F}, \mathcal{C}_\mathrm{init}, \hat{\mathcal{C}})$, where $\mathcal{I}$ is the issue description, $\mathcal{F}$ is the fault location, $\mathcal{C}_\mathrm{init}$ is the initial retrieved context, and $\hat{\mathcal{C}}$ is the minimal sufficient context identified by OCD. We transform this corpus into a pointwise classification dataset of $(q, s, y)$ triples, where $q$ is a structured query, $s$ is a code segment, and $y \in \{0, 1\}$ indicates whether the oracle retained that segment. Each initial context $\mathcal{C}_\mathrm{init}$ is parsed into a set of code segments $\{s_1, \ldots, s_n\}$.

We use this corpus to fine-tune Qwen3-Reranker-0.6B and obtain \projName, a lightweight cross-encoder that defines a scoring function $f(q, s; \theta) \in [0,1]$. Given a structured query $q$ built from the issue description and fault location, with a code segment $s$ from the initial context, the model predicts whether $s$ should be retained to preserve patch sufficiency. A reranker model is well suited to this setting because the supervision is defined at the segment level, and inference only requires scoring a fixed pool of candidate segments rather than generating context from scratch.

We employ Low-Rank Adaptation (LoRA) for parameter-efficient fine-tuning. This reduces trainable parameters by over an order of magnitude, making fine-tuning feasible on commodity hardware while preserving the base model's code understanding. We use LoRA with rank $r=16$ and scaling factor $\alpha=32$, applied to the query, key, value, and output projection matrices. Training runs for 3 epochs with an effective batch size of 16 (per-device batch size 4 with 4 gradient accumulation steps), using AdamW with learning rate $2 \times 10^{-4}$, 10\% linear warmup, weight decay of 0.01, and gradient clipping at norm 1.0. We use bfloat16 mixed precision with gradient checkpointing to fit training within commodity GPU memory. The best checkpoint is selected by AUC-ROC on a held-out validation set comprising 10\% of instances.

A key challenge in our dataset is that the oracle discards far more code segments than it retains---a typical instance has a compression ratio well below 50\%, meaning negatives substantially outnumber positives. Without correction, the model would learn to trivially predict ``discard'' for all segments. We address this through two complementary mechanisms: (1) a class-level weight that upscales the loss contribution of positive (retained) segments in proportion to the imbalance ratio, ensuring that missing a necessary segment is penalized more heavily than retaining a redundant one; and (2) the per-sample role-aware weights, which further modulate individual examples based on their structural informativeness. Together, these mechanisms aim to encode that it is worse to lose a patch-critical dependency than to include a few extra lines of supporting code.

At inference time, given a new instance, the system first parses $\mathcal{C}_\mathrm{init}$ into segments and constructs the structured query $q$ as in training. Each segment is scored by the cross-encoder in batches. For segments that exceed the model's context window, a sliding-window strategy scores overlapping windows independently and aggregates them, ensuring that long functions or class bodies are evaluated fairly rather than truncated. Finally, the system assembles the compressed context through budget-aware greedy selection. The token budget is controlled by a configurable compression rate.

\section{Evaluation}

\begin{table*}[t]
\centering
\small
\setlength{\tabcolsep}{3pt}
\caption{Comparison of context compression methods on SWE-bench Verified (500 instances). For each downstream LLM, the number in parentheses denotes the mean initial context length in tokens before compression. Compression is reported as the reduction rate relative to the initial context length (higher is more aggressive). \textbf{Bold} denotes the best result per column.}
\label{tab:compression_results}
\resizebox{\textwidth}{!}{
\begin{tabular}{lccc ccc ccc}
\toprule
\multirow{2}{*}{Method}
& \multicolumn{3}{c}{DeepSeek-V3.2 (15,921 tokens)}
& \multicolumn{3}{c}{Qwen3-Coder-Next (7,761 tokens)}
& \multicolumn{3}{c}{GPT-5.2 (15,690 tokens)} \\
\cmidrule(lr){2-4} \cmidrule(lr){5-7} \cmidrule(lr){8-10}
& Resolved & Rate & Compression
& Resolved & Rate & Compression
& Resolved & Rate & Compression \\
\midrule
\texttt{No\_compression} & 239 & 0.478 & 1.00$\times$
                         & 200 & 0.400 & 1.00$\times$
                         & 266 & 0.532 & 1.00$\times$ \\
\texttt{No\_context}     & 231 & 0.462 & --
                         & 193 & 0.386 & --
                         & 262 & 0.524 & -- \\
\texttt{LLMLingua-2}       & 245 & 0.490 & 2.98$\times$
                         & 193 & 0.386 & 2.80$\times$
                         & 272 & 0.544 & 2.94$\times$ \\
\texttt{LongCodeZip}     & 238 & 0.476 & 5.02$\times$
                         & 199 & 0.398 & 7.40$\times$
                         & 284 & 0.568 & 7.67$\times$ \\
\texttt{SWE-Pruner}       & 233 & 0.466 & 4.10$\times$
                         & 192 & 0.384 & 49.75$\times$
                         & 270 & 0.540 & 22.37$\times$ \\
\texttt{SWEzze}      & \textbf{261} & \textbf{0.522} & 6.03$\times$
                         & \textbf{210} & \textbf{0.420} & 5.95$\times$
                         & \textbf{289} & \textbf{0.578} & 6.55$\times$ \\
\bottomrule
\end{tabular}
}
\end{table*}



Our evaluation answers the following research questions:

\begin{enumerate}
    \item[\textbf{RQ1}] How does \projName compare to existing compression baselines in terms of compression ratio and  efficiency?

    \item[\textbf{RQ2}] Does \projName preserve or improve downstream issue resolution rates compared to baselines across different LLMs?

    \item[\textbf{RQ3}] What compression-related root causes lead to failed issue resolution?
\end{enumerate}

\paragraph*{Dataset} We conduct our evaluation on SWE-bench Verified~\cite{jimenez2023swe}, a curated subset of 500 real-world GitHub issues spanning 12 popular Python repositories. Each instance consists of an issue description, the repository snapshot at the time of filing, a ground-truth patch, and a regression test suite. SWE-bench Verified is widely adopted as a standard benchmark for LLM-based issue resolution, providing a controlled yet realistic testbed where issues range from single-line fixes to multi-file refactorings.

We obtain the code context for each instance using Agentless~\cite{xia2024agentless}. Agentless hierarchically localizes an issue by first identifying relevant files, then classes and functions within those files, and finally pinpointing specific code elements. During this process, it retrieves a ranked list of related elements at each granularity. We use the top-$10$ strategy to select elements from the ranked list, chosen to match the context size used in prior compression work~\cite{shi2025longcodezip}; since the total cost of our experiments already exceeds 2800\$, we did not perform additional ablation over this parameter. After Agentless determines suspicious locations, we treat the edit targets as the focal point and retain the remaining top-ranked related elements as the supporting context.


\paragraph*{Baselines} We compare \projName against three compression baselines spanning distinct compression paradigms:

\begin{itemize}
    \item \textbf{LLMLingua-2}~: A token-level compression method that trains a compact encoder to predict per-token preservation probabilities via data distillation. 
    \item \textbf{LongCodeZip}~: A hierarchical code-specific compressor that ranks functions by conditional perplexity and prunes blocks. 
    \item \textbf{SWE-Pruner}~:  A task-aware code compressor that uses a lightweight reranker to estimate token relevance and prunes lines whose aggregated scores fall below a threshold.
    \item \textbf{No Compression}: The full evaluation context is passed directly to the repair model without any reduction.
    \item \textbf{No Context}: Only the issue description and fault location are provided, with no supporting code context. 
\end{itemize}

To ensure fairness, we configure each baseline at its recommended or best-aligned operating point. For LLMLingua-2 and SWE-Pruner, we adopt default parameter settings specified in their respective publications. For LongCodeZip, we derive the target compression ratio from the effective compression levels reported in LongCodeZip's original experiments and set the compression rate to 5$\times$. For \projName, we also set the compression rate to 5$\times$.



\paragraph*{Models} To assess generalizability of \projName across different LLMs, we evaluate each compression method under three downstream repair models that differ in architecture, scale, and provider: GPT-5.2~\cite{singh2025openai}, DeepSeek-V3.2~\cite{liu2024deepseek}, and Qwen3-Coder-Next~\cite{cao2026qwen3}.


\paragraph*{Evaluation Metrics} We assess compression methods along four dimensions that jointly capture the cost effectiveness trade-off of context compression:

\begin{itemize}
    \item \textbf{Resolution Rate}: The fraction of instances for which the repair model generates a patch that passes the full regression test suite. This is the primary effectiveness metric, directly measuring whether compression preserves the semantic content necessary for correct patch synthesis.
    \item \textbf{Compression Rate}: The ratio $|\mathcal{C}_\mathrm{init}| /|\hat{\mathcal{C}}| $ measured in tokens, where high values indicate more aggressive compression.  This metric quantifies the reduction in context size.
    \item \textbf{Token Count}: The total number of tokens consumed during the repair step, comprising both compressed prompt tokens and completion tokens. 
    \item \textbf{Compression Time}: The wall-clock time required to compress a single instance, measuring the computational overhead that the compression step introduces into the repair pipeline.
\end{itemize}



\begin{figure*}[t]
    \centering
    \includegraphics[width=\textwidth]{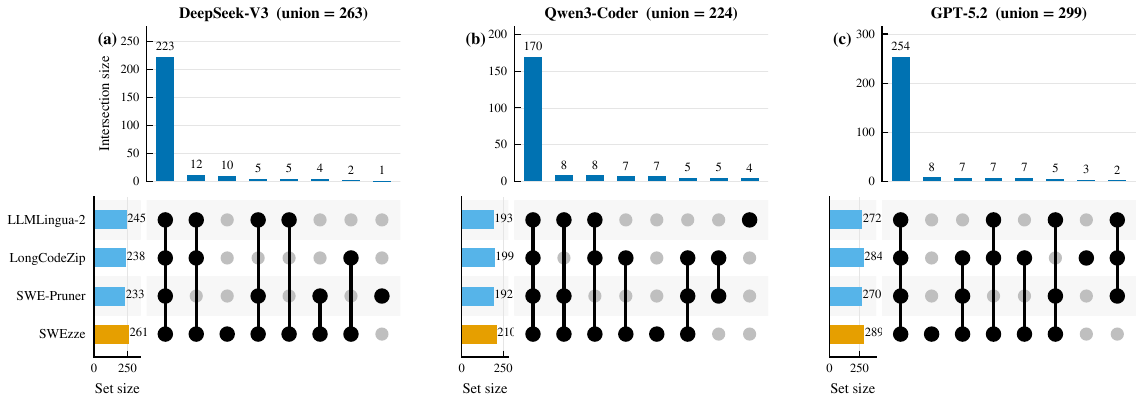}
    \vspace{-4mm}    
    \caption{UpSet plots of resolved-instance overlaps across compression methods on SWE-bench Verified. Each panel corresponds to one downstream repair model. Top bars show the size of resolved-instance intersections, while the left bars show the total number of resolved instances per method, with \projName highlighted in orange. The plots reveal both the shared wins and the additional instances uniquely recovered by \projName.\label{fig:upset}}
    \vspace{-2mm}
\end{figure*}

\subsection{RQ1: Compression Efficiency and Cost}
\label{sec:eval:rq1}

\begin{table}[t]
\centering
\small
\setlength{\tabcolsep}{4pt}
\caption{Prompt and total token consumption per instance across downstream LLMs. \projName substantially reduces the end-to-end token cost relative to the uncompressed setting while retaining sufficient context for repair.}
\label{tab:token_cost}
\resizebox{\columnwidth}{!}{
\begin{tabular}{
l
S[table-format=5.0] S[table-format=5.0]
S[table-format=4.0] S[table-format=5.0]
S[table-format=5.0] S[table-format=5.0]
}
\toprule
\multirow{2}{*}{Method}
& \multicolumn{2}{c}{DeepSeek-V3.2}
& \multicolumn{2}{c}{Qwen3-Coder-Next}
& \multicolumn{2}{c}{GPT-5.2} \\
\cmidrule(lr){2-3} \cmidrule(lr){4-5} \cmidrule(lr){6-7}
& {Prompt $\downarrow$} & {Total $\downarrow$}
& {Prompt $\downarrow$} & {Total $\downarrow$}
& {Prompt $\downarrow$} & {Total $\downarrow$} \\
\midrule
\texttt{No\_compression} & 21808 & 23194 & 9457 & 13901 & 16893 & 17408 \\
\texttt{No\_context}     &  6359 &  7627 & 1675 &  4394 &  2255 &  2767 \\
\texttt{LLMLingua-2}     & 11480 & 12864 & 4408 &  7236 &  7237 &  7731 \\
\texttt{LongCodeZip}     &  9177 & 10569 & 2658 &  6028 &  4166 &  4667 \\
\texttt{SWE-Pruner}      &  9834 & 11214 & 1753 &  4822 &  2912 &  3393 \\
\texttt{\projName}       &  8658 & 10230 & 2922 &  6707 &  4492 &  4992 \\
\midrule
\multicolumn{1}{r}{\texttt{\projName} vs. \texttt{No\_compression}}
& \multicolumn{1}{c}{$-60.3\%$} & \multicolumn{1}{c}{$-55.9\%$}
& \multicolumn{1}{c}{$-69.1\%$} & \multicolumn{1}{c}{$-51.8\%$}
& \multicolumn{1}{c}{$-73.4\%$} & \multicolumn{1}{c}{$-71.3\%$} \\
\bottomrule
\end{tabular}
}
\end{table}

\begin{table}[t]
\centering
\small
\setlength{\tabcolsep}{6pt}
\caption{Mean compression latency per instance (seconds) for each method across the three LLMs. Lower is faster.}
\label{tab:compression_latency}
\resizebox{\columnwidth}{!}{
\begin{tabular}{lccc}
\toprule
Method & DeepSeek-V3.2 & Qwen3-Coder-Next & GPT-5.2 \\
\midrule
\texttt{LLMLingua-2}   & 0.9  & 0.44 & 0.9 \\
\texttt{LongCodeZip} & 34.6 & 10.9 & 27.8 \\
\texttt{SWE-Pruner}   & 1.2  & 0.48 & 1.1 \\
\texttt{SWEzze}      & 10.0 & 2.5  & 5.9 \\
\bottomrule
\end{tabular}
}
\vspace{-1em}
\end{table}

\Cref{tab:compression_results} shows that across all three LLMs \projName's compression rate stays close to 6$\times$ (6.03$\times$ on DeepSeek-V3.2, 5.95$\times$ on Qwen3-Coder-Next, and 6.55$\times$ on GPT-5.2). This pattern is more stable than that of the baselines. LLMLingua-2 is conservative, remaining below 3$\times$ compression in every setting. LongCodeZip compresses more, but its compression rate varies more across models. SWE-Pruner's compression rate ranges from 4.10$\times$ on DeepSeek-V3.2 to 49.75$\times$ on Qwen3-Coder-Next. The weak result of SWE-Pruner on DeepSeek-V3.2 can be explained by the fact that the issue description and fault location consume much of SWE-Pruner's hint-generation budget. The achieved compression rate may exceed the target rate because the target defines only a minimum compression level. Once no budget remains for another code segment, or no remaining segment meets the selection threshold, the method stops adding segments.

Higher compression rate leads to token cost savings. Compared with \texttt{No\_compression}, \projName removes 13,150 prompt tokens per instance on DeepSeek-V3.2, 6,535 on Qwen3-Coder-Next, and 12,401 on GPT-5.2, which corresponds to prompt reductions of 60.3\%, 69.1\%, and 73.4\%, respectively. The end-to-end token budget, measured as the sum of prompt and completion tokens, still falls by 55.9\%, 51.8\%, and 71.3\% relative to the uncompressed setting. Compared with LLMLingua-2, \projName further reduces the total token budget by 20.5\% on DeepSeek-V3.2 and 35.4\% on GPT-5.2. On Qwen3-Coder-Next, however, its total cost remains higher than that of the more aggressive LongCodeZip and SWE-Pruner. It shows that \projName does not optimize only for the smallest possible token cost, but aims for more balanced, controlled compression.

Averaged over the three LLMs, LLMLingua-2 and SWE-Pruner are the fastest methods, requiring only 0.75 and 0.93 seconds per instance, respectively, while LongCodeZip is much slower at 24.4 seconds. \projName lowers this overhead to 6.1 seconds on average, making it about 4.0$\times$ faster than LongCodeZip. The same pattern appears across all backends: relative to LongCodeZip, \projName reduces compression latency by 71.1\% on DeepSeek-V3.2, 77.1\% on Qwen3-Coder-Next, and 78.8\% on GPT-5.2. At the same time, it is 8.2$\times$ slower than LLMLingua-2 and 6.6$\times$ slower than SWE-Pruner, which reflects the extra cost of structured segmentation and reasoning on compression queries. While \projName is not the cheapest compressor in absolute terms, it offers a better balance among compression rate, stability, and latency than the baselines.

\begin{tcolorbox}[colback=gray!5, colframe=black!20, arc=2pt, boxrule=0.3mm, breakable, sharp corners, left=2pt, right=2pt, top=2pt, bottom=2pt]
\textbf{RQ1:} \projName achieves stable $\sim$6$\times$ compression across LLMs, reduces the total token count by 51.8\%--71.3\% over \texttt{No\_compression}; it is 4.0$\times$ faster than LongCodeZip on average.
\end{tcolorbox}

\subsection{RQ2: End-to-End Issue Resolution}
\label{sec:eval:rq2}

\Cref{tab:compression_results} shows that \projName achieves the best issue resolution performance across all three LLMs.  Its resolution rate reaches 0.522 on DeepSeek-V3.2, 0.420 on Qwen3-Coder-Next, and 0.578 on GPT-5.2. Relative to No\_compression, these results correspond to absolute gains of 4.4, 2.0, and 4.6 percentage points (relative improvements of 9.2\%, 5.0\%, and 8.6\%). When compared with No\_context, \projName's gains are 6.0, 3.4, and 5.4 percentage points (13.0\%, 8.8\%, and 10.3\% in relative improvement). This pattern suggests that repository context helps issue resolution only when intelligently selected. The much smaller gap between No\_compression and No\_context further suggests that a substantial part of the raw retrieved context is not necessary and may even hinder the repair process.

Importantly, \projName's improvements do not simply come from solving the same easy instances as the competing compressors. \Cref{fig:upset} shows that the largest intersection in every panel is the four-way overlap, with 223 instances on DeepSeek-V3.2, 170 on Qwen3-Coder-Next, and 254 on GPT-5.2. This indicates that all methods solve a large shared set of relatively easy instances. Even so, \projName still covers the largest resolved set on every comparison. Compared with the strongest baseline for each model, this adds 16 instances on DeepSeek-V3, 11 on Qwen3-Coder, and 5 on GPT-5.2, which corresponds to relative gains of 6.5\%, 5.5\%, and 1.8\%. Measured against the union of all resolved instances, the coverage of \projName reaches 99.2\% on DeepSeek-V3, 93.8\% on Qwen3-Coder, and 96.7\% on GPT-5.2. This means that the advantage of \projName is not limited to higher overall accuracy. It also recovers a broader set of instances, including cases missed by other compressors.

\begin{figure}
    \centering
    \includegraphics[width=0.45\textwidth]{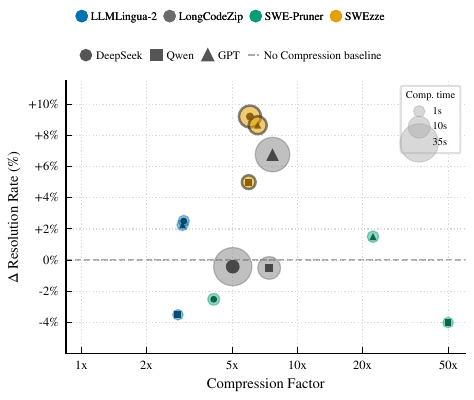}
    \caption{Tradeoff between compression rate, resolution rate improvement over
  the no-compression baseline ($\Delta = 0\%$, dashed), and compression latency
  (bubble area) across four methods and three LLMs (marker shape).
  \textsc{\projName} (orange) achieves the highest resolution-rate gains
  ($+5.0\%$--$+9.2\%$) at a moderate compression rate (${\sim}6\times$) and
  latency.}
    \label{fig:tradeoff}
      \vspace{-1em}
\end{figure}

As shown in \Cref{fig:tradeoff}, the baselines tend to fall into less favorable operating regions. LLMLingua-2 stays in a low-cost but low-gain region, offering only marginal improvement or even causing degradation. SWE-Pruner represents the other extreme. It can apply very aggressive compression, but its downstream behavior is unstable, suggesting that excessive pruning often removes information needed for repair. LongCodeZip is more competitive in compression strength, but its gains are less consistent and come with substantially higher latency. In contrast, \projName consistently achieves the largest improvements in resolution rate, while maintaining a moderate compression rate of about 6$\times$ and a practical runtime overhead. These results support the central claim of this work: for repository-level issue resolution, the goal is not to maximize compression aggressiveness, but to preserve the small subset of context that is truly sufficient for generating a correct patch.

\begin{tcolorbox}[colback=gray!5, colframe=black!20, arc=2pt, boxrule=0.3mm, breakable, sharp corners, left=2pt, right=2pt, top=2pt, bottom=2pt]
\textbf{RQ2:} \projName achieves the best issue resolution performance across all three LLMs, improving resolution rates by 5.0–13.0\% to the uncompressed and no-context settings. It also covers the broadest set of resolved instances, reaching 93.8\%--99.2\% of the union of all solved cases.
\end{tcolorbox}

\begin{figure}[t]
  \begin{center}
    \includegraphics[width=0.9\columnwidth]{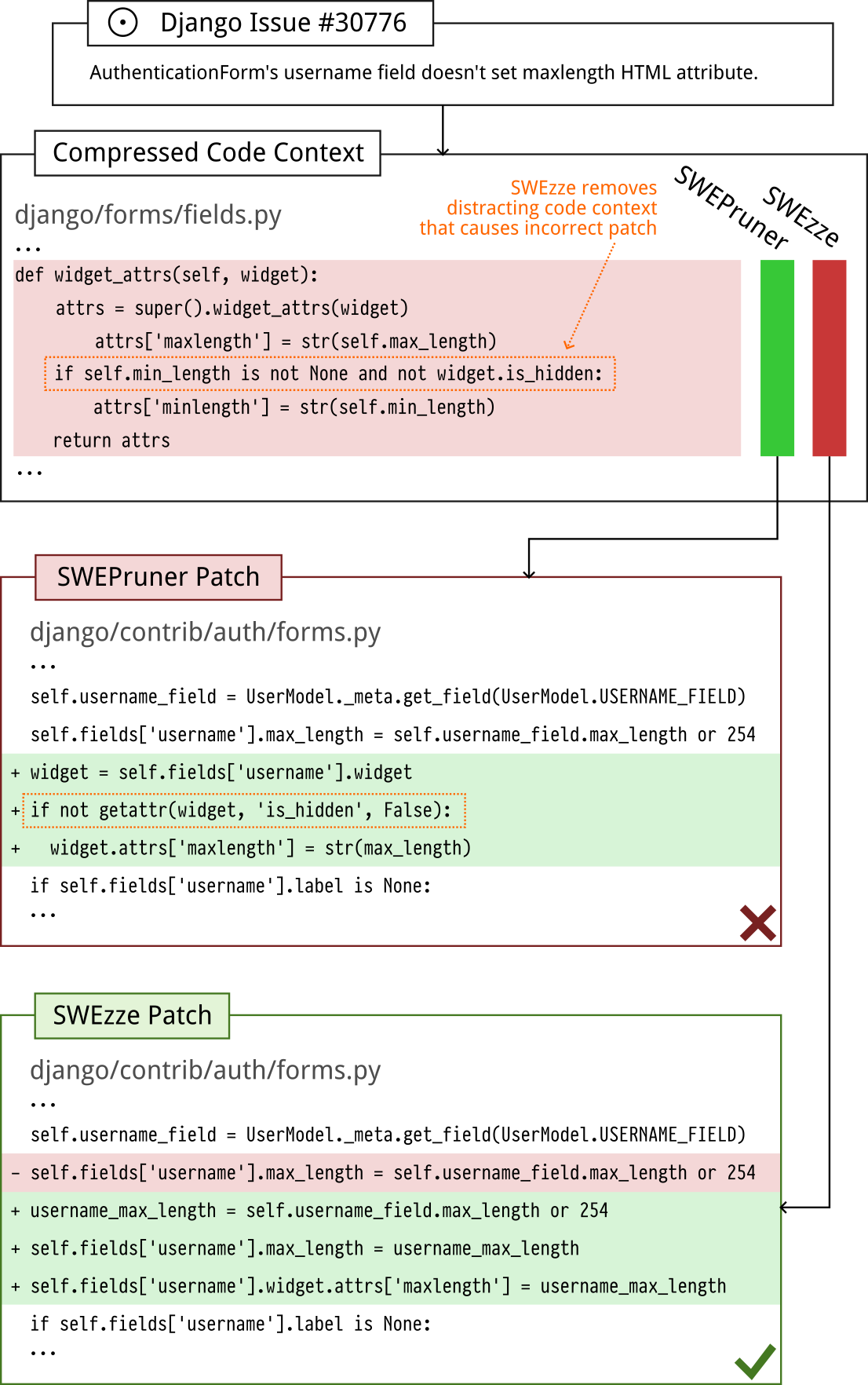}
  \end{center}
  \vspace{-2mm}
  \caption{SWE-Pruner retains a distracting fragment that leads to an incorrect patch, whereas \projName removes it, enabling correct resolution. The compressed context uses the same visual notation as \Cref{fig:context_comparison}.\label{fig:distraction}}
  \vspace{-4mm}
\end{figure}

\subsection{RQ3: Analysis of Compression Failures}



Based on the UpSet plot, we manually inspected unresolved instances, and found three recurring types of failures. First, a compressor may retain distracting code, which pulls the LLM toward irrelevant logic. Second, the model may not score code segments with enough precision. This problem often occurred in projects with overridden functions or multiple implementations that share the same signature and have similar function bodies. In such cases, small code differences may lead to divergent behaviors, but the model cannot distinguish them reliably. Third, overly aggressive compression may damage the syntactic structure of the original code and produce fragmented contexts that obscure semantic relations. We uniformly sampled 30 instances unresolved by GPT-5.2 for which the uncompressed code context produced a passing patch. Among these instances, 5 are due to distracting context, 8 to imprecise segment scoring, and 17 to disrupted syntactic structure.

Figure~\ref{fig:distraction} shows a representative case where retained context distracts the repair model. The issue requires assigning the \texttt{maxlength} attribute to the \texttt{username} field. In this case, \projName, \texttt{No\_context}, and \texttt{No\_compression} all resolve the issue successfully, whereas LongCodeZip and SWE-Pruner fail. The failure arises because these two methods, especially SWE-Pruner, discard most of the surrounding context and keep only the \texttt{widget\_attrs} function. This function contains a predicate that checks whether \texttt{widget.is\_hidden} is false, and one branch assigns a value to \texttt{widget.max\_length}. With only this compressed context, the repair model generates a patch that adds an unnecessary constraint on \texttt{is\_hidden}, which causes the tests to fail. This example shows that context compression can hurt repair performance not only by removing essential information, but also by keeping misleading fragments that divert the model from the true repair logic.

\begin{tcolorbox}[colback=gray!5, colframe=black!20, arc=2pt, boxrule=0.3mm, breakable, sharp corners, left=2pt, right=2pt, top=2pt, bottom=2pt]
\textbf{RQ3}: Manual analysis shows that most failures are due to distracting retained context, insufficiently precise semantic discrimination between similar code segments, and breaking syntactic structure through overly aggressive compression.
\end{tcolorbox}

\section{Threats To Validity \& Discussion}
\label{sec:data_discussion}

To assess whether \projName generalizes across model families, we evaluate with three issue-resolution models from distinct providers and architectures: GPT-5.2, DeepSeek-V3.2, and Qwen3-Coder-Next. Crucially, none of these is the model used during OCD data construction (Qwen3-Coder-30b), which means the compressed contexts were distilled under one model yet consistently benefit different ones. This cross-model transfer suggests that the fix ingredients extracted by OCD capture genuine semantic dependencies of the issue rather than artifacts of a particular model's decoding strategy, and that the resulting compression is largely model-agnostic.

The search process relies on ground-truth patches and test coverage information that are unavailable at inference time. This creates a distribution gap between training and deployment: the model must learn to approximate the oracle's judgments using only information available from the issue description and retrieved code. We address this gap in the model training phase by formulating the task as learning to predict which units the oracle would retain. Importantly, the inputs required by OCD---(issue, patch, test suite) triples---are naturally produced by any software project that uses pull requests with continuous integration. This means OCD training data can in principle be mined from any repository with a sufficiently rich merge history, not only from curated benchmarks.

Our search explicitly targets minimal sufficient contexts rather than merely sufficient ones. This design choice reflects the hypothesis that minimal contexts provide cleaner supervision signals---they contain less noise and more directly capture the causal dependencies required for repair. However, minimality is defined with respect to the specific LLM and generation parameters used during data construction; the truly minimal context may vary across models.



\section{Related Work}

\paragraph*{LLM Context} Previous research noticed that LLMs struggle with processing long contexts~\cite{liu2024lost,li2024loogle}, raising the so-called ``lost-in-the-middle'' problem. Worse, Shi et al~\cite{shi2023large} noticed that LLMs tend to be distracted by irrelevant context, which significantly degrades performance. Although recent work on long context handling partially alleviated this problem~\cite{zhang2025recursive}, state-of-the-art LLMs continue to exhibit this weakness.

\paragraph*{Context Retrieval} To resolve issues in real-world repositories, an LLM requires relevant code context containing necessary project-specific dependencies. FitRepair~\cite{xia2023plastic} uses hard-coded heuristics to augment prompts with contextual code data. LLM agents employ more flexible mechanisms such as embedding-based retrieval~\cite{xia2024agentless}, retrieval-augmented generation (RAG)~\cite{lewis2020retrieval} such as matching relevant code snippets with BM25~\cite{robertson2009probabilistic}, or agentic retrieval that invokes code search APIs~\cite{zhang2024autocoderover}. Because these techniques rely on syntactic similarity, and capture only shallow semantic properties, they typically overapproximate the necessary context, which result in decreased generation efficiency and increased cost.

\paragraph*{Prompt Compression} Prompt compression aims to reduce the length of prompts to improve the efficiency of processing LLM inputs~\cite{li2025prompt,wan2023efficient}. A large body of works focus on task-agnostic natural language compression, such as LlmLingua~\cite{jiang2023llmlingua} and LlmLingua2~\cite{pan2024llmlingua}. Techniques such as TACO-RL~\cite{shandilya2025taco} offer task-specific natural language compression, that optimizes prompt for a specific task. Despite their practicality, such approaches are ineffective in selecting relevant code context for issue resolution, because source code demands strict syntactic integrity and requires taking into account structural dependencies.

\paragraph*{LLM-based Issue Resolution \& Program Repair} LLMs enabled effective automated program repair~\cite{le2019automated} that can resolve issues in real-world repositories, with representative techniques including ChatRepair~\cite{xia2024automated}, SWE-Agent~\cite{DBLP:conf/nips/YangJWLYNP24}, and AutoCodeRover~\cite{zhang2024autocoderover}. In this work we rely on Agentless~\cite{xia2024agentless} because it represents a canonical issue resolution workflow, which is widely used to evaluate frontier LLMs on issue resolution~\cite{deng2025swe}.

\paragraph*{Code Compression \& Information Selection} Early works such as Maniple~\cite{parasaram2024fact} selected relevant code fragments for LLM-based program repair using a random forest, however this approach is not fine-grained enough to identify precise fix ingredients. Xiao et al.~\cite{xiao2025improving} suggested to reduce agent trajectories by removing useless, redundant, and expired information to reduces computational costs. In contrast to \projName, this approach is not code-specific. LongCodeZip~\cite{shi2025longcodezip} introduced a code-specific compression framework; it first ranks function-level chunks using conditional perplexity to align with the task, and then applies fine-grained block-level pruning via knapsack algorithm. Similarly, SWE-Pruner~\cite{wang2026swe} fine-tunes a lightweight reranker to score each token's task-conditioned relevance, and prunes lines whose aggregated scores fall below a threshold. Their key limitation is that the ranking criteria, perplexity or relevance score, do not address the problem of retaining ingredients for issue resolution, since they merely measure relatedness of code fragment to the bug. In contrast, \projName reframes the problem from compressing for relevance to distilling for sufficiency. By leveraging oracle feedback, we move beyond statistical proxies to explicitly search for the minimal context configuration that enables the generation of a correct patch.

\section{Conclusion}

We presented a code context compression framework for LLM-based issue resolution. Unlike prior compressors that rely on perplexity or lexical similarity, \projName is trained on minimal sufficient contexts produced by Oracle-guided Code Distillation (OCD), a two-phase search combining genetic algorithms and hierarchical delta debugging. OCD identifies the smallest context subsequence that still enables a correct patch, thus capturing ingredients required for issue resolution rather than surface-level relatedness. Evaluated on SWE-bench Verified with three frontier LLMs, \projName achieves stable compression while also improving resolution rates over previous compression baselines.



\section*{Acknowledgments}

We thank Yue Pan for assistance with the implementation, and Yihan Dai, Haotian Xu, and Dimitrios Stamatios Bouras for their comments on an early draft.

\bibliographystyle{ACM-Reference-Format}
\bibliography{ref}

\end{document}